\documentclass[apl,article, superscriptaddress,reprint]{revtex4-1}

\usepackage{graphicx}
\usepackage{epsfig}
\usepackage{color}
\usepackage[hypertex,unicode=true]{hyperref}

\hypersetup{
        colorlinks=false,
        citecolor=blue,
        linkcolor=black,
        urlcolor=blue}

\newcommand{\halfl}{\ensuremath{{\scriptstyle \frac{1}{2}}}}

\newcommand{\un}   [1]{\ensuremath{\,\mathrm{#1}}}
\newcommand{\unit} [1]{\un{#1}}
\newcommand{\unitt}[1]{\ensuremath { \mathrm {#1} }}
\newcommand{\intd}    {\ensuremath{\,\mathrm{d }}}

\newcommand{\pder} [2]{\ensuremath{\frac{\partial #1}{\partial #2}}}
\newcommand{\pderl}[2]{\ensuremath{\partial #1/\partial #2}}

\newcommand{\tderl}[2]{\ensuremath{\mathrm{d} #1/\mathrm{d} #2}}

\newcommand{\neff} {\ensuremath {n_{\mathrm{eff}}}}
\newcommand{\dopt} {\ensuremath {d_{\mathrm{opt}}}}
\newcommand{\delec} {\ensuremath {d_{\mathrm{elec}}}}

\begin{document}

\title{Broadband nanoelectromechanical phase shifting of light on a chip}

\author{M.~Poot}
\email{menno.poot@yale.edu} \affiliation{Department of
Electrical Engineering, Yale University, New Haven, CT 06520,
USA}

\author{H.~X.~Tang}
\email{hong.tang@yale.edu} \affiliation{Department of
Electrical Engineering, Yale University, New Haven, CT 06520,
USA}

\date{\today}

\begin{abstract}
We demonstrate an optomechanical phase shifter. By
electrostatically deflecting the nanofabricated mechanical
structure, the effective index of a nearby waveguide is changed
and the resulting phase shift is measured using an integrated
Mach-Zehnder interferometer. Comparing to thermo-optical phase
shifters, our device does not consume power in static operation
and also it can operate over large frequency, wavelength, and
power ranges. Operation in the MHz range and sub-$\mu$s pulses
are demonstrated.
\end{abstract}

\maketitle
\newpage
Integrated on-chip optics has many advantages over free-space
optics, such as compactness, scalability, and stability. This
is especially important for quantum computation with light
\cite{knill_nature_quantum_computation_linear_optics}. We
envision a structure with on-chip sources of correlated photons
\cite{davanco_APL_onchip_heralded,
jiang_arxiv_integrated_correlated_photons,
laucht_PRX_integrated_QD_source}, that are guided through
nanophotonic structures implementing a quantum-computation
algorithm, then storing \cite{wang_PRL_storage} the result in a
quantum optomechanical resonator
\cite{poot_physrep_quantum_regime}, finally followed by
detection with highly efficient integrated superconducting
single photon detectors \cite{pernice_natcomm_SSPD,
schuck_APL_OTDR, schuck_SR_NbTiN_SSPD_dark_count}. In such a
platform all operations are performed on a single chip,
reducing coupling losses, enhancing the device stability, and
making the entire system easily scalable.
\begin{figure}[b]
\includegraphics{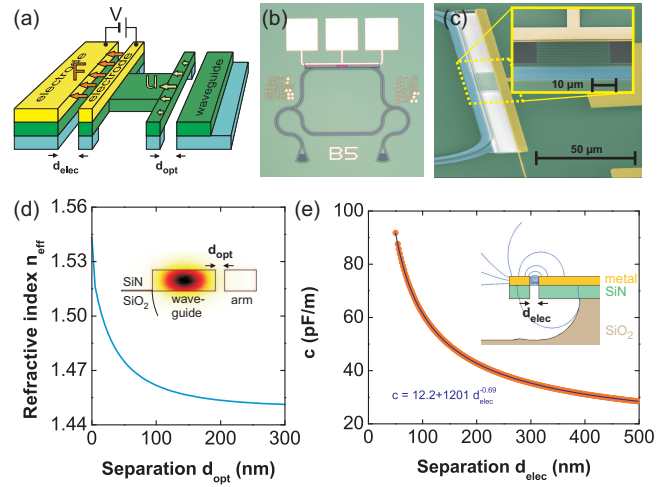}
\caption{(a) Schematic operation of the opto-electromechanical
phase shifter. A voltage $V$ applied between the electrodes
creates a force on the left side of the H-resonator which
results in an in-plane displacement $u$ away from the waveguide.
The change in refractive index leads to a shifted phase of the
light traveling through the waveguide.
(b) Optical micrograph of a full device showing the waveguides
of the MZI connecting to grating couplers, the mechanical
resonator, and the gold contact pads. The two outer pads are
grounds, connected through the metalized arms of the mechanical
resonator; the voltage is applied to fixed electrode via the
center pad. (c) Colorized scanning-electron micrograph of a
resonator under a $70^\mathrm{o}$ angle. The inset shows a zoom
of the central region of the resonator with the photonic
crystal block and the gap between the two electrodes (yellow)
and between the resonator and the waveguide (green). (d)
Simulated separation-dependence of $\neff$ at $\lambda_0 = 1550
\un{nm}$. The inset shows the calculated optical mode profile
for $\dopt = 150 \un{nm}$. (e) Simulated dependence of the
capacitance per unit length on the electrode gap (symbols) and
a power-law fit (line). The inset shows a cross-section with
the electric field lines for $\delec = 225 \unit{nm}$.
\label{fig:overview}}
\end{figure}

Phase shifters are an essential part for the quantum
information processing with photons as well as for other
applications. These are devices that can controllably change
the phase of the light field. When a photon travels a length
$\ell$ through a material with a refractive index $n$, the
phase that it acquires is $\phi = 2\pi n \ell / \lambda_0$,
where $\lambda_0$ is the free-space wavelength. Since the path
length $\ell$ is difficult to change in an integrated optical
circuit, phase shifts are induced by changing the refractive
index. Current on-chip phase modulators use different
techniques to achieve this: First of all, one can change the
effective refractive index of a waveguide by injecting carriers
into it \cite{xu_nature_integrated_EOM,
reed_natphot_Si_modulators}. This technology is widely used in
the semiconductor and photonic industry, but cannot be used
with important insulating materials such as SiN. Another
commonly used way to phase-modulate light on a chip is to use
the temperature dependence of the refractive index: by placing
a heater close to the waveguide, the refractive index can be
modulated \cite{vlasov_nature_slow_light,
pruessner_OE_thermooptic_tuning, beggs_SPIE_slow_light}.
However, this technique typically requires a large power to be
dissipated on the chip \cite{song_OE_thermooptic} making it
challenging to implement at the cryogenic temperatures required
for superconducting detectors. Furthermore, even at room
temperature, thermal crosstalk between closely spaced
thermo-optical devices is a severe problem. Other designs for
phase modulators have employed movable parts to change the
resonant wavelength of an on-chip cavity
\cite{winger_OE_optoelectomechanical, sun_APL_SHF_opto_NEMS,
miao_NJP_MEMS_disk, sridaran_OE_Si_disk_MEMS,
abdulla_OE_racetrack_tuning}. However, all these devices are
intrinsically narrowband in nature, either to the wavelength of
the input light, or to the modulation frequency when resonant
modulation is required.

Here we demonstrate broadband on-chip phase shifting of
near-infrared light by electrostatically actuating a
nanomechanical structure that is placed in the vicinity of an
optical waveguide. The device (Fig. \ref{fig:overview}(a)-(c))
consists of SiN photonic circuits and a mechanical resonator
that can be displaced by applying a voltage between two
electrodes. The in-plane deflection modifies the effective
refractive index of the waveguide, thereby phase-shifting the
light traveling through it \cite{guo_APL_phase_shifter}. By
integrating the waveguide in one of the arms of a Mach-Zehnder
interferometer (MZI) this phase shift is detected. Note again
that our devices do not rely on light within a narrow
wavelength range in contrast to cavity-based systems: In
principle, the operation wavelength is only limited by the
bandwidth of the waveguides used (the cut-off wavelength of the
waveguide is around 1800 nm), but in our current implementation
the wavelength range is set by the transmission window of the
grating couplers spanning up to 50 nm in the telecom wavelength
range.
\begin{figure*}[tb!]
\includegraphics{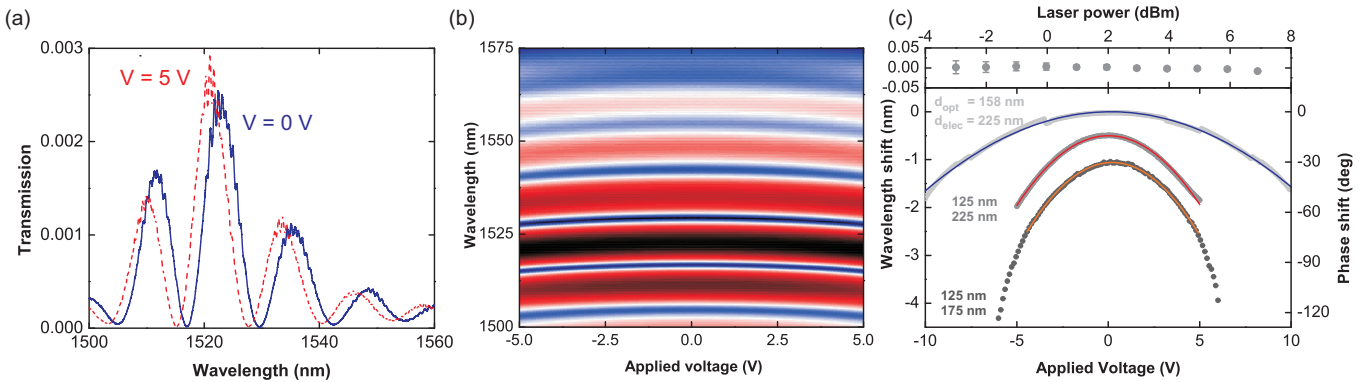}
\caption{(a) Transmission spectra of the MZI for $V = 0
\unit{V}$ (solid) and for  $V = 5 \unit{V}$ (dashed). (b)
Colormap of the wavelength dependence of the transmission
versus the applied voltage: high (low) transmission is dark red
(blue). (c) $\Delta \lambda(P_L)$ (top) and $\Delta \lambda(V)$
(bottom) for three different devices (symbols) and a quadratic
fit (solid lines). The curves have been offset for clarity and
the phase shift on the right axis is calculated using a period
of $12.6 \un{nm}$. All periods deviate less than 10\% from this
mean value. \label{fig:static}}
\end{figure*}

The mechanical part of the device shown in Fig.
\ref{fig:overview}(a)-(c) consists of four SiN arms with width
$w_a = 500 \un{nm}$ and length $L_a = 100\unit{\mu m}$. One
side of the arms is fixed at the clamping points and the other
side is connected to a block at the center of the resonator,
forming an ``H'' (Fig. \ref{fig:overview}). An electrode runs
over one pair of arms and is separated from a second, fixed,
electrode by a small gap $\delec$. There is also a narrow gap
$\dopt$ between the remaining two arms and the waveguide. To
facilitate the release and in order to prevent leakage of light
from the waveguide into the block, holes are etched in the
block forming a photonic crystal with a bandgap around the
operating wavelength yet retaining the rigidity of the block.
Still, the most important point of our ``H-resonator'' design
is that the electrodes are far away from the waveguide thus
avoiding absorption of the light in the metal
\cite{davanco_OE_slot_mode_design}.


The phase shifters are made on a silicon handle substrate with
$3.3 \unit{\mu m}$ oxide (cladding) and a $t=330 \unit{nm}$
thick (device) layer of high-stress SiN ($n = 2.0$ and stress
$\sigma \sim 700 \unit{MPa}$) on top. The devices are
fabricated in a three-step electron beam lithography process,
followed by a wet release. The electrodes (5 nm Cr and 210 Au)
are patterned first. Then the couplers and waveguides are
written and dry-etched. The nitride is not completely etched: a
thin ($\sim 30 \unit{nm}$) layer of SiN remains on the silicon
oxide to protect the latter during the wet release. Next, the
moveable structures are defined and etched all the way through
the SiN. Both etching steps are aligned w.r.t. the electrodes
using metal alignment markers defined in the first step. Note
that the last lithography step has to be aligned within this
small gap between the electrodes. Typically, the alignment
between all layers is better than 40 nm.

The effect of the in-plane motion of the mechanical structure
on the propagation through the waveguide is calculated with
two-dimensional finite-element (FEM) simulations in COMSOL.
Figure \ref{fig:overview}(d) shows the dependence of the
effective refractive index $\neff$ of the first TE-like mode on
the separation $\dopt$ between the 1000-nm-wide waveguide and
the arms of the resonator. Without the resonator (i.e., for
$\dopt \rightarrow \infty$) $\neff$ is 1.450 but increases to
1.462 at $\dopt = 100 \un{nm}$. Thus by displacing the
resonator the phase acquired by the light traveling through the
upper arm can be controlled. The total phase shift accumulated
over the entire interaction length of the device $L_i = 170
\un{\mu m}$ (i.e., the length over which the waveguide runs
close to the resonator; $L$ is the length of the whole device)
is then:
\begin{equation}
\Delta \phi \approx \frac{2\pi}{\lambda_0}\int_{-L_i/2}^{L_i/2} \pder{\neff}{u} u(x) \intd x \approx \frac{2\pi L_i}{\lambda_0}\pder{\neff}{\dopt} \bar u, \label{eq:phase}
\end{equation}
where $\bar u$ is the average in-plane displacement along the
arms closest to the waveguide. To attain the largest phase
shifts the derivative has to be as large as possible and hence
$\dopt$ should be as small as possible. Devices with $\dopt$
between 125 and 225 nm and $\delec = 175$ and 225 nm have been
fabricated; for narrower gaps the devices tend to stick to
either the waveguide or the fixed electrode. The capacitance
between the electrodes depends strongly on $\delec$ as shown in
Fig. \ref{fig:overview}(e). By applying a voltage $V$ an
attractive electrostatic force per unit length $F/L = -\halfl
V^2 \pderl{c}{\delec}$ results, where $c$ is the capacitance
per unit length \cite{poot_physrep_quantum_regime}. For a
capacitive gap of 225 nm the data in Fig. \ref{fig:overview}(e)
yields a total force of $1.0 \unit{\mu N}$ on the resonator at
$V = 10 \un{V}$. The mechanical compliance of the structure is
mainly determined by the four arms so that the spring constant
$k \approx 4k_a = 4\sigma w_a t/L_a = 4.6 \un{N/m} $ if the
force were localized on the block instead of being distributed
along the entire H-resonator. The displacement that is obtained
with this simple estimate is 200 nm; a full FEM simulation of
the static displacement shows that the block displaces 134 nm
towards the electrode and that the average displacement is
$\bar u = 75 \un{nm}$ away from the waveguide. Combining
everything shows than the expected phase shift is $\Delta \phi
\sim - 1.0 \pi$ for a device with $\dopt = 125 \un{nm}$.

In the experiment, light from a tunable laser source with power
$P_L$ is sent from a single mode optical fiber onto the chip
using a grating coupler (the triangular structures in Fig.
\ref{fig:overview}(b)) and passed though the MZI which converts
the phase shift into an amplitude change. The modulated light
is coupled into a second fiber and detected with a
photodetector.
AC and DC electrical signals are combined with a bias-T and are
applied to the device using a high-frequency
ground-signal-ground probe. Figure \ref{fig:static}(a) shows
the optical transmission of the device. MZI fringes with a
period of $12.6 \unit{nm}$ are imprinted on the overall
transmission profile of the grating couplers.
Now when a voltage is applied, the fringes of the MZI
interferometer shift towards shorter wavelengths (Fig.
\ref{fig:static}(a)). This can be understood as follows: the
voltage pulls the resonators away from the waveguide, reducing
$\neff$ and thus decreasing the phase accumulated in the
(90-$\un{\mu m}$ longer) upper arm of the MZI. To bring the
phase difference back to its original value, a smaller
wavelength is needed and hence the fringes shift to the left.
Figure \ref{fig:static}(b) shows a colormap of the transmission
of the device for different applied voltages. For both positive
and negative polarities the curves bend downward indicating
that the electrostatic force is always attractive
\cite{poot_physrep_quantum_regime}. By extracting the position
of the fringes, the tuning curves $\Delta \lambda(V)$ shown in
Fig. \ref{fig:static}(c) are obtained. As discussed above,
devices with smaller gaps $\dopt$ and $\delec$ are expected to
have the largest wavelength shifts. Indeed, the fringes of the
device with $\dopt = 125 \un{nm}$ and $\delec = 175 \un{nm}$
shift more than 3 nm at -6 V. A quadratic fit of $\Delta
\lambda(V)$ yields a curvature of $-133 \un{pm/V^2}$. This is
much larger than the curvature reported in Ref.
\cite{sun_APL_SHF_opto_NEMS} and comparable to the narrowband
device of Ref. \cite{winger_OE_optoelectomechanical}. Using the
period of the MZI fringes, the wavelength shift can be
converted back to a phase shift. In this case $\pderl{^2\Delta
\phi}{V^2} = -60 \un{mrad/V^2}$ which can be compared to the
curvature calculated using Eq. (\ref{eq:phase}) $\pder{^2\Delta
\phi}{V^2} = -\frac{2 \pi L_i L }{\lambda_0 \overline{k}}
\pder{\neff}{\dopt}\pder{c}{\delec} \approx -174
\un{mrad/V^2}$, where $\bar k = \pderl{\bar u}{F}$. The
calculated curvature is a bit higher than the measured one,
possibly due to the unknown stress in the metal and the
difference between the actual and designed gaps on which the
derivatives depend sensitively. In any case, the right axis of
Fig. \ref{fig:static}(c) shows that a single device generates a
phase shift of more than $90^\mathrm{o}$ and a full $\pi$ phase
shift would be achieved at $V_\pi\sim 10 \un{V}$. Larger phase
shifts (or, equivalently, a smaller $V_\pi$) can easily be
obtained by cascading multiple H-modulators.

\begin{figure}[tb]
\includegraphics{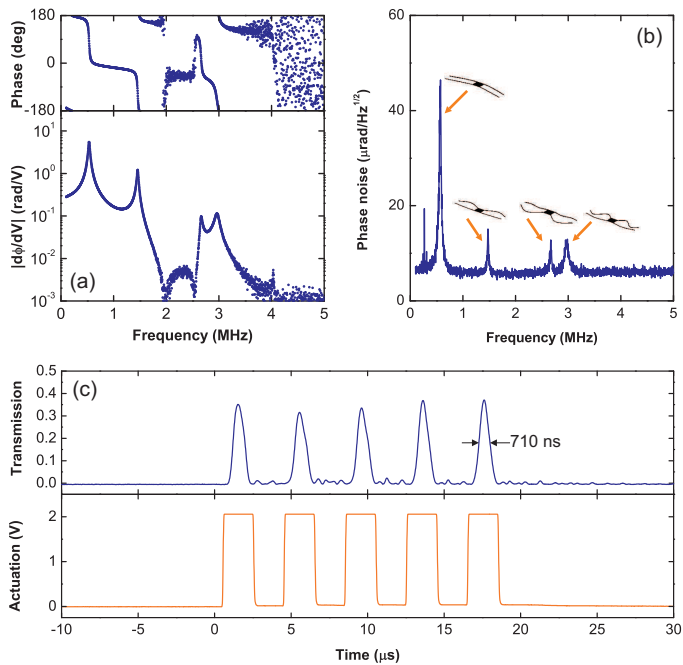}%
\caption{
(a) Magnitude (bottom) and phase (top) of the frequency
response of phase shifter with $\dopt = 125 \un{nm}$ and $\delec = 225
\un{nm}$ as measured with a network analyzer at a static voltage $V = 5 \unit{V}$.
(b) The phase noise spectrum of the same modulator at $V = 5 \unit{V}$.
Peaks due to the mechanical eigenmodes appear on top of a
background due to the laser noise and the insets show the
corresponding in-plane mechanical modes.
(c) Fast phase pulses. The wavelength is set to the bottom of one
of the MZI fringes and a train of five actuation pulses (bottom)
is applied. The phase changes on a sub-$\unitt{\mu s}$ time
scale as indicated by the full-width-at-half-maximum value of the
last peak. The transmission is normalized by the peak transmission
of the MZI. \label{fig:dynamic}}
\end{figure}

Figure \ref{fig:dynamic}(a) shows the dynamic behavior of the
device. The output of a network analyzer is added to the static
voltage and the response at the excitation frequency is
measured. These measurements show that our phase shifter is not
only broadband in the optical sense, but operates over a large
modulation frequency range: The 3 dB point occurs at 1.0 MHz
and coherent operation up to a few MHz is possible before the
signal is lost in the noise. The magnitude $|\tderl{\phi}{V}|$
contains a series of peaks. These are the eigenfrequencies of
the H-resonator where the dynamic phase shift is resonantly
enhanced by the mechanical quality factor $Q_m$. A flatter
frequency response can be engineered by lowering $Q_m$, for
example by operating in liquid or by coating with mechanically
lossy polymers. Alternatively, one could utilize negative
feedback to operate the device at lower quality factors
\cite{miao_NJP_MEMS_disk}.

By measuring the output of the photodetector with a spectrum
analyzer and converting the voltage noise power spectral
density back to a phase, the noise spectrum of Fig.
\ref{fig:dynamic}(b) is obtained. It has a background of $6
\un{\mu rad/Hz^{1/2}}$ which is due to the noise of the laser
and of the photodetector; these are thus not intrinsic to the
device. However, the peaks are the Brownian motion of the
eigenmodes (the peak at 0.27 MHz is, however, due to the
laser). At the fundamental mode (at 0.58 MHz) the entire
resonator is moving, whereas the higher modes are mainly
located on the arms. The noise performance of the phase shifter
also can be improved by intentionally reducing the mechanical
quality factor.

The device can also generate fast phase shifts as demonstrated
in Fig. \ref{fig:dynamic}(c). Here, a pulse train of five
rectangular pulses is applied \cite{liu_natnano_timedomain}
(bottom) resulting in the output signal in the top panel. Due
to the high resonance frequency of the H-resonator sub-$\mu s$
pulses are realized. The pulse width of $710 \un{ns}$ is
already much shorter than state-of-the-art thermo-optic devices
\cite{song_OE_thermooptic, atabaki_OE_thermooptic}, but by
engineering the pulse shape, even shorter pulses are possible
\cite{harjanne_IEEE_thermooptic_pulse}.

Finally, the dynamic range of the optical power is considered.
At low optical power $P$, the device does not feel the
influence of the light traveling through the waveguide, but at
high power the attractive optical gradient force will pull the
resonator closer to it. This was actually the operation
principle of the devices studied in Ref.
\cite{fong_OE_tunable_coupler} and was for example used to
build an optomechanical radio-frequency amplifier
\cite{li_natcomm_switch_amplification}, but is unwanted for an
electrically controlled phase modulator. The gradient force
\cite{povinelli_OL_gradient_force} per unit length is $F_G/L =
-n_g P / n v_0 \cdot \pderl{n}{u}$ where $v_0$ is the speed of
light in vacuum. The power in the waveguide that causes the
same force as $V = 1 \un{V}$ is about 70 mW for the parameters
discussed above; for a 10 times higher operation voltage this
value changes to 7 W, indicating that in practice the optical
power does not play a role. Indeed, in the experiment no change
of the fringe position with laser power was observed up to 6.9
dBm (Fig. \ref{fig:static}(c)). Besides being broadband in the
electrical and optical sense, our optomechanical phase shifter
is thus also capable of handling a large range of optical
power.

\begin{acknowledgments}
M.P thanks Netherlands Organization for Scientific Research
(NWO) / Marie Curie Cofund Action for providing a Rubicon
postdoctoral fellowship. This work was partly funded by the
DARPA/MTO ORCHID program through a grant from AFOSR. H.X.T.
acknowledges support from a Packard Fellowship in Science and
Engineering and a career award from National Science
Foundation. Facilities used were supported by Yale Institute
for Nanoscience and Quantum Engineering and NSF MRSEC DMR
1119826. We thank Wolfram Pernice for discussions.
\end{acknowledgments}



\end{document}